\PassOptionsToPackage{unicode}{hyperref}
\PassOptionsToPackage{hyphens}{url}
\documentclass[
  11pt,
]{article}
\usepackage{xcolor}
\usepackage[margin=1in]{geometry}
\usepackage{amsmath,amssymb}
\usepackage{iftex}
\ifPDFTeX
  \usepackage[T1]{fontenc}
  \usepackage[utf8]{inputenc}
  \usepackage{textcomp} % provide euro and other symbols
\else % if luatex or xetex
  \usepackage{unicode-math} % this also loads fontspec
  \defaultfontfeatures{Scale=MatchLowercase}
  \defaultfontfeatures[\rmfamily]{Ligatures=TeX,Scale=1}
\fi
\usepackage{lmodern}
\ifPDFTeX\else
\fi
\IfFileExists{upquote.sty}{\usepackage{upquote}}{}
\IfFileExists{microtype.sty}{% use microtype if available
  \usepackage[]{microtype}
  \UseMicrotypeSet[protrusion]{basicmath} % disable protrusion for tt fonts
}{}
\makeatletter
\@ifundefined{KOMAClassName}{% if non-KOMA class
  \IfFileExists{parskip.sty}{%
    \usepackage{parskip}
  }{% else
    \setlength{\parindent}{0pt}
    \setlength{\parskip}{6pt plus 2pt minus 1pt}}
}{% if KOMA class
  \KOMAoptions{parskip=half}}
\makeatother
\NewDocumentCommand\citeproctext{}{}

\makeatletter
 \let\@cite@ofmt\@firstofone
 \def\@biblabel#1{}
 \def\@cite#1#2{{#1\if@tempswa , #2\fi}}
\makeatother
\newlength{\cslhangindent}
\newlength{\csllabelwidth}
\newenvironment{CSLReferences}[2] % #1 hanging-indent, #2 entry-spacing
 {\begin{list}{}{%
  \setlength{\itemindent}{0pt}
  \setlength{\leftmargin}{0pt}
  \setlength{\parsep}{0pt}
  \ifodd #1
   \setlength{\leftmargin}{\cslhangindent}
   \setlength{\itemindent}{-1\cslhangindent}
  \fi
  \setlength{\itemsep}{#2\baselineskip}}}
 {\end{list}}
\usepackage{calc}

\usepackage{bookmark}
\IfFileExists{xurl.sty}{\usepackage{xurl}}{} % add URL line breaks if available
\makeatletter
\@ifundefined{xmpquote}{}{}
\makeatother
\hypersetup{
  pdftitle={Coherence{,} charity and triangulation in statistical modelling},
  pdfauthor={David J. T. Sumpter},
  hidelinks,
  pdfcreator={LaTeX via pandoc}}

\title{Coherence, charity and triangulation in statistical modelling}
\author{David J. T. Sumpter}
\date{August 2026}

\begin{document}
\maketitle

Department of Information Technology, Uppsala University,
david.sumpter@it.uu.se

\section{Abstract}\label{abstract}

Bayesian statistics rests on a few familiar distinctions: frequentist
vs.~Bayesian, objective versus subjective probability, a model versus
the data it is fitted to, a prior versus a posterior. Here, I use Donald
Davidson's ``third dogma of empiricism'' to critique such distinctions
in terms of scheme-content dualisms. With a single running example ---
estimating how often a football (soccer) team scores --- I argue that a
degree of belief is in itself objective, that model and data are better
seen as parts of one belief system than as different kinds of things,
and that Bayes' theorem describes relations within that system rather
than data updating a model from outside. What matters instead is how
belief systems are built, checked and rebuilt. To this end, I use
Davidson's constructive programme of radical interpretation to suggest
that a belief system should be not only coherent (as statisticians
already require) but also charitable and triangulated --- answerable to
other people and to a shared world. I argue that these three constraints
are exactly what good statistical practice --- from eliciting priors to
checking models --- already tries to satisfy and talking about them
explicitly could improve this practice.

\section{1. Introduction}\label{introduction}

A commonly used modelling formalism is to write \[
\mathrm{P}(M \mid D)
\] as the probability of a model \(M\) given a particular data set
\(D\). We can view \(\mathrm{P}(M \mid D)\) as our degree of belief in a
model: the probability we think \(M\) is true given the data \(D\)
available to us (Edwards et al. 1963; Etz and Vandekerckhove 2018). In
practice, \(\mathrm{P}(M \mid D)\) can be thought of in terms of the
betting odds we would be prepared to accept on the model being correct.
We say our beliefs are coherent if they don't lead us to accept a
collection of bets on which we will lose. Coherent beliefs obey the
usual axioms of probability (Ramsey 1931; {de Finetti} 1937; Savage
1954).

This so-called Bayesian approach is particularly useful in parameter
estimation and it also allows different models to be compared. For
example, in a study of how groups of glass prawns interact, competing
models --- each with its own parameters \(\theta\) --- were compared
(Mann et al. 2013). The winning model, the one considered to offer the
best explanation of the data, was the one for which \(\theta\) maximised
the posterior \(\mathrm{P}(\theta \mid D)\). Similar approaches have
been used for parameter estimation and model selection in models of
democratic development (Spaiser et al. 2014), cosmology (Trotta 2008),
the inference of evolutionary trees (Ronquist and Huelsenbeck 2003), the
epidemiology of COVID-19 ({Flaxman et al.} 2020), medical
decision-making and the synthesis of clinical evidence (Spiegelhalter
and Best 2003), and in capturing human cognition (Tenenbaum et al.
2011).

The Bayesian framework is said to supply a rule for revising our beliefs
about models as new data arrives: Bayesian updating (Gelman et al. 2013;
Howson and Urbach 2006). We start from a prior, \(\mathrm{P}(M)\) ---
our degree of belief in a model before seeing the data. The model itself
supplies a likelihood, \(\mathrm{P}(D \mid M)\), the probability it
attaches to the data. Bayes' theorem then combines the two into a
posterior,
\[\mathrm{P}(M \mid D) = \frac{\mathrm{P}(D \mid M)\,\mathrm{P}(M)}{\mathrm{P}(D)},\]
our revised degree of belief once the data are in. Beliefs can be
refined step by step as evidence accumulates. In this sense, Bayesian
updating lets us learn from data.

There is evidence that humans who take a Bayesian-like updating approach
to their decision-making make better decisions (Schoemaker and Tetlock
2016). In the Good Judgment Project --- a tournament in which thousands
of volunteers forecast geopolitical events --- the most accurate
forecasters, the so-called superforecasters, were set apart not by
access to privileged information but by how they handled probabilities
(Tetlock and Gardner 2015; Mellers et al. 2015). They expressed their
beliefs on a fine-grained probability scale and revised them in the
manner Bayes' rule prescribes: through many small updates as evidence
arrived, rather than occasional dramatic swings (Atanasov et al. 2020).

While Bayesian reasoning has proved effective in practice, a long line
of criticism questions whether the Bayesian picture can carry the weight
placed on it. One objection concerns the priors: if a probability is
meant to express ignorance, there seems to be no coherent way to choose
it (Norton 2008). A second concerns the reasoner: the framework assumes
logical omniscience, requiring an agent to assign a probability to
everything they believe, which no real person does (Hacking 1967).
Another problem arises because, in any realistic modelling problem,
there are infinitely many models we could write down. So the uncertainty
that matters most is often not over the parameters of a given model, but
over which model to consider in the first place (Spiegelhalter and Best
2003).

A deeper concern lies with the very idea of Bayesian updating. Sprenger
argues that the conditional probability \(\mathrm{P}(D \mid M)\) is
routinely glossed as updating our degree of belief, yet the number a
model assigns to the data is fixed internally when the model is set up
(Sprenger 2020). Gelman and Shalizi provide multiple applied examples,
from social sciences and economics, where working Bayesian analysis
proceeds not by inductive updates, but by large scale revision of
beliefs (Gelman and Shalizi 2013). They emphasise a more
falsification-led approach to building models, checking them against the
data, and rebuilding them when they fail.

The approach I take here is complementary to the above and starts from
Donald Davidson's work on, what he called, the third dogma of
empiricism. Expanding on Quine's critique of the analytic--synthetic
distinction (1951), Davidson identifies a dualism of scheme and content
--- an organising conceptual scheme on one side, and uninterpreted
empirical content, the ``given'', waiting to be organised on the other
(Davidson 1974). Much of the richest work in statistics has been framed
around divisions (or the dissolution of divisions) of just this shape
--- frequentist versus Bayesian (Efron 1986; Mayo 2018); objective
versus subjective Bayes (Berger 2006; Goldstein 2006); the question of
how models interact with the data they are fitted to (Suppes 1962;
Bokulich 2020); the idea of a true, data-generating model that inference
converges upon, which Box denied (1976); and the question of whether a
parameter names a real feature of the world or an artefact of the model
(Borsboom et al. 2003). Davidson's claim is that any such dualisms are
unintelligible: we can make no clear sense of a content wholly
independent of the scheme that organises it.

Davidson's critique of the scheme and content distinction was not
entirely new. Writing in 1868 Peirce denied that we possess any
cognition not determined by a previous cognition, and so any
uninterpreted given at all (Peirce 1868). At the start of the 20th
Century, James and Dewey set themselves against divisions between an
experience passively given and a mind that comes along to organise it
(James 1907; Dewey 1929), anticipating Davidson's approach (Rorty 1986,
1979; Godfrey-Smith 2002). Nor is Davidson's critique the most
up-to-date. Ramsey, de Finetti and Savage all give strong arguments for
a subjective account of beliefs (Ramsey 1931; {de Finetti} 1937; Savage
1954), which takes us some way towards Davidson's conclusions. That data
is shaped by modelling is widely recognised (Suppes 1962; Edwards 2010;
Bokulich 2020). For example, Bokulich shows how models used for
filtering, interpolating and scaling (to give three of seven examples
provided in the article) shape data collection (Bokulich 2020).

What marks Davidson out is his constructive programme of radical
interpretation (1973). On this programme we insist that a belief system
obeys three constraints, which he named at different points across his
work rather than together in one place (Davidson 1980, 1991). The first
is \emph{coherence}: those degrees of belief must hang together
consistently, obeying the axioms of probability (although for Davidson
it was logic rather than probability) --- the constraint the subjective
tradition already recognises. The second is \emph{charity}: we interpret
a person so that they come out, on the whole, rational and largely
right. The third is \emph{triangulation}: what a person's beliefs are
about is fixed between them, the others they must be intelligible to,
and the world all of them inhabit.

Rationality is often identified with only one of these constraints,
namely coherence, together with expected-utility maximisation and
Bayesian updating (Ramsey 1931; {de Finetti} 1937; Neumann and
Morgenstern 1944; Savage 1954). Coherence defines the rational agent of
economics and artificial intelligence (Russell and Norvig 2021), the
Bayesian reasoner of cognitive science (Oaksford and Chater 2007;
Griffiths et al. 2010; Tenenbaum et al. 2011), and the inferential brain
of computational neuroscience (Friston 2010; Clark 2013); even
challenges from bounded and ecological rationality are framed against
this ideal (Simon 1955; Gigerenzer and Selten 2001). In these settings,
rationality is an internal property of an individual agent. As we will
see, Davidson's framework suggests that a focus on coherence alone
misses important components of rationality and effects how we should
design belief systems. It is charity and triangulation that keep a
belief system from collapsing into mere subjectivity even though it is,
unmistakably, unique to every individual.

I start by setting out the argument for probability as a degree of
belief, following the subjective tradition of Ramsey, de Finetti and
Savage(1931; 1937; 1954), emphasising differences to Davidson's
non-subjective approach. I then look at the relationship between models
and data, following the philosophy of data literature in emphasising
that there is no clean separation between the two (Suppes 1962; Edwards
2010; Bokulich 2020). Then I show that Bayes' theorem is a relation
internal to the belief system, so that any story of data arriving to
update a model is incorrect. Here, like Gelman and Shalizi, I relocate
Bayesian reasoning away from induction and towards the building and
checking of models on many levels(Gelman and Shalizi 2013). I close the
paper by proposing a more radical view: that rational approaches should
continually revise their entire belief system, using all three of
Davidson's constraints: charity, coherence and triangulation.

Several of the conclusions I draw here are already established. That
observation is theory-laden was argued long ago (Hanson 1958; Kuhn
1962), and the models-as-mediators tradition has shown in detail how
models stand between theory and data (Morgan and Morrison 1999; Suppes
1962). That objectivity in science is not a view from nowhere but
something achieved socially is the lesson of a large literature in
social epistemology (Longino 1990; Douglas 2009). And the foundations of
Bayesian inference have been examined at length (Sprenger and Hartmann
2019; Howson and Urbach 2006; Mayo 2018). What is new here is using
Davidson's radical interpretation to dissolve the dualisms around which
probabilistic thinking is often described, to turn Bayes' theorem into a
relation internal to the belief system, and to reveal that best practice
in applied statistics is well described as in terms of his charity and
triangulation constraints (in addition to the coherence which is usually
emphasised by statisticians).

\section{2. A probabilistic belief
system}\label{a-probabilistic-belief-system}

\subsection{2.1 Definition}\label{definition}

We start by defining a probability space and random variables. A
probability space consists of three components: \(\Omega\) which is the
set of all possible \textbf{outcomes}; an event space \(\mathcal{F}\)
which is a collection of subsets of \(\Omega\), whose elements are
called \textbf{events}. \(\mathrm{P}\) is a \textbf{probability
measure}, that is, a function \(\mathrm{P} : \mathcal{F} \to [0,1]\)
which assigns a probability. In addition to the probability space, we
will also define \textbf{random variables} which describe how we see
events in terms of numbers. Here and elsewhere, I avoid giving technical
definitions (for these see, for example, (Grimmett and Stirzaker 2020))
and instead give an example to illustrate these four elements. The
purpose of this section is to relate outcomes, events, the probability
measure and random variables to a belief system.

For events and outcomes this mapping is straightforward. Imagine that
you are watching a football (soccer, as it is referred to in the USA)
match. And each time the team you support shoots, you note down if it
was a ``goal (\(G\))'' or a ``not a goal (\(N\))''. For three shots, all
possible outcomes are \[
\Omega = \{ GGG, GGN, GNG, NGG, GNN, NGN, NNG, NNN \}
\] A subset of events might be, for example, all the sequences where
there was exactly one goal (\(\{ GNN, NGN, NNG \}\)) or all the
sequences where the first shot was a goal and the second and third shots
were either a goal or not, (\(\{ GNN, GGN, GNG, GGG \}\)). All in all
there are 256 potential events for the three shots, (including the set
\(\Omega\) of all outcomes and the empty set \(\emptyset\)). Notice that
it is not a single shot that is an event here, but statements about all
shots -- such as ``there was exactly one goal''
(i.e.~\(\{ GNN, NGN, NNG \}\)); ``the first shot was a goal''
(i.e.~\(\{ GNN, GGN, GNG, GGG \}\)); and ``each shot was either a goal
or not'' (i.e.~\(\Omega\)).

Random variables are also uncontroversial. They turn events into
numbers. For example, we can write \(D_t=1\) if shot \(t\) is a goal and
\(D_t=0\) if shot \(t\) it isn't a goal. We can also keep track of the
number of goals in the match. So, for example, \[
D_1=0, D_2=0, D_3=1, G_1=0, G_2=0, G_3=1,
\] describes three shots, of which the third is a goal, with the \(D_t\)
taking care of shots and \(G_t\) the number of goals scored at that
point in time. There are different possible random variables which can
be constructed from events. For example, tracking the number of goals
scored after each shot was a choice, and in this case \(G_3=1\)
corresponds to the event \(\{ GNN, NGN, NNG \}\).

The next step is to assign probabilities to random variables. If we
already know that the probability that a shot is a goal is 1 in 10 or
\(0.1\) and that shots are independent (unaffected by each other) then
\[
\mathbb{P}(D_3=1) = 1/10 \tag{1}
\] We write this with a blackboard \(\mathbb{P}\) on purpose: in this
framing the number is treated as a chance the world carries, directly
assigned to an outcome --- ``the probability that a goal is scored on
the third shot (i.e.~\(D_3=1\))''. It is the reading we will end up
setting aside. Everywhere else we write \(\mathrm{P}\), for a
probability read as our own degree of belief.

There is another way of creating a probability function using these
observations. Imagine we introduce a fourth shot into our probability
space and we look at the event, ``there was exactly one goal in the
first three shots (and there was either a goal or not on the fourth
shot)'' (i.e.~\(\{ GNNG, NGNG, NGGG, GNNN, NGNN, NGGN \}\). In this
situation we know that \(G_3=1\). In this case, we might assume that the
best estimate we have is the frequency of goals up to this point. This
is given by \(G_3/3\), the average number of goals per shot. We can then
write that \[
\mathrm{P}(D_4=1 \mid G_3=1) = 1/3
\] The role of \(G_3\) in this case is to summarise what we have seen so
far and help us predict the future.

In general, we can write this prediction as \[
\mathrm{P}(D_{t+1}=1 \mid G_{t}=g) = g/t \tag{2}
\] The value \(g/t\) is a parameter in a model of the data collected so
far. Notice that this approach changes how we should read the
probability compared to \(\mathbb{P}\) in equation (1). There the
probability was given to us and we said it was ``the probability that a
goal is scored on the third shot''. But now
\(\mathrm{P}(D_{t+1}=1 \mid G_{t}=g)\) is ``the degree to which we
believe that a goal will be scored on the \(t\)'th shot given the
outcome of the \(t\) earlier shots''.

There are some practical problems with the model proposed in equation
(2). We might have been lucky to see a goal with the third shot, and one
goal in three was an over-representation of goals. A clear example of
how our model can be wrong can be seen after the second shot, where we
would write \[
\mathrm{P}(D_3=1 \mid \{D_1=0, D_2=0\}) = \mathrm{P}(D_3=1 \mid G_2=0) = 0/2 = 0
\] Our model predicts that shot 3 won't be a goal under any
circumstances, but clearly it is! The problem is that we have adopted a
100\% certainty in our beliefs about goals after only a small number of
observations.

To account for this, we can introduce a new random variable, which we
will call \(\Theta\). This variable represents our estimate of the
probability of a goal. In equation (2) we simply set \(\Theta=g/t\), but
now we are going to say that \(\Theta\) is a random variable which takes
value \(\theta\) with probability
\(\mathrm{P}(\Theta =\theta \mid G_t=g)\). This means that, instead of
specifying that \(\Theta\) takes a particular value, we specify (through
\(\mathrm{P}(\Theta =\theta \mid G_t=g)\)) that \(\Theta\) can take one
of a wide range of values. Now, ``the probability with which we believe
that a goal is scored on the \(t+1\)'th shot given our estimate of the
probability of a goal is \(\theta\)'' is written as
\(\mathrm{P}(D_{t+1}=1 \mid \Theta=\theta)\).

We now have two expressions involving \(\Theta\), namely
\(\mathrm{P}(\Theta \mid G_t=g)\) and
\(\mathrm{P}(D_{t+1}=1 \mid \Theta=\theta)\). All random variables
should obey the axioms of probability. I haven't detailed those axioms
here (and again refer the interested reader to a text on probability)
but one of the results that can be proven true from those axioms is
Bayes' theorem. Namely, that for any two random variables, the
relationship \[
\mathrm{P}(X \mid Y) = \frac{\mathrm{P}(Y \mid X ) \mathrm{P}(Y)}{\mathrm{P}(X)}
\] holds. Another definition arising from probability is that of
conditional independence between variables. Variables \(X\) and \(Z\)
are said to be conditionally independent after the outcome of \(Y\), if
we can write that \[
\mathrm{P}(X \mid Y, Z) = \mathrm{P}(X \mid Z)
\] Conditional independence holds for \(D_{t+1}\) and \(G_t\) given the
outcome of \(\Theta\). We can thus write, \[
\mathrm{P}(D_{t+1}=1 \mid \Theta=\theta, G_t=g) = \mathrm{P}(D_{t+1}=1 \mid \Theta=\theta)
\] In words, this says that while we need to know the number of goals to
make our estimate of \(\Theta\), once we know our estimate \(\Theta\) we
no longer need to know the number of goals to predict if the next shot
will be a goal (i.e.~\(D_{t+1}\)). One way to think of this is as
follows: if I tell you that I think the next shot will be a goal with
probability \(13.4\)\%, you don't need to know how many shots and goals
I have actually seen in producing that estimate. (You might want to
challenge my assumption, but that is another question. For now, we are
only concerned with the value of my estimate, not how I came up with
it).

Let's now put our three variables into Bayes' theorem, \[
\mathrm{P}(\Theta=\theta \mid D_{t+1}=1, G_t=g) = \frac{\mathrm{P}(D_{t+1}=1 \mid \Theta=\theta, G_t=g) \mathrm{P}(\Theta=\theta \mid G_t=g)}{\mathrm{P}(D_{t+1}=1 \mid G_t=g)}
\] Now, using conditional independence, we can write \[
\mathrm{P}(\Theta=\theta \mid D_{t+1}=1, G_t=g) = \frac{\mathrm{P}(D_{t+1}=1 \mid \Theta= \theta) \mathrm{P}(\Theta=\theta \mid G_t=g)}{\mathrm{P}(D_{t+1}=1 )} 
\] This version of Bayes' theorem is a coherence constraint between
random variables we have set up based on the axioms of the probability
space.

Now we have set up the belief system, we now look at how we talk about
it.

\subsection{2.2 Degrees of belief}\label{degrees-of-belief}

We have looked at two alternative ways of talking about probabilities.
The first, which we discussed briefly and labelled \(\mathbb{P}\),
treats the probability that the next shot will be a goal as a feature of
the shot itself --- a physical chance, or a long-run frequency. This is
what is usually meant by an objective probability (Reiss and Sprenger
2020).

The other, which we discussed in more detail and settled on, labels
\(\mathrm{P}\) our degree of belief that the next shot will be a goal.
This is what is usually called a subjective, or personal, probability
(Savage 1954). The word ``subjective'' is easily misread: it does not
mean the number is arbitrary or just what we happen to think (Sprenger
2018). It underlies our beliefs accumulated by a reasonable or rational
analysis of the data. For example, while we don't see
\(\mathrm{P}(D_{t+1}=1)\) as the objective probability a shot will be a
goal, it is an average over our uncertainty about the goal-rate, i.e.~
\[\mathrm{P}(D_{t+1}=1) = \int \theta\,\mathrm{P}(\Theta=\theta)\,\mathrm{d}\theta = \mathbb{E}[\Theta]\]
It is a summary of our own belief distribution, built from the counts we
have seen and the prior we brought. It is fixed both by the belief
system we have set up and the data which we have seen.

It might appear natural to ask whether, as data accumulate, our belief
\(\mathrm{P}\) converges to some objective probability, like
\(\mathbb{P}\). There are mathematical results showing that two
observers whose priors agree on what is possible will see their
predictions merge (Blackwell and Dubins 1962) (although not always
(Diaconis and Freedman 1986)). However, what these observers settle on
is the parameter the model and the data pick out together. It is their
beliefs which converge to be the same.

The argument that \(\mathrm{P}\) measures only degrees of belief, and is
unrelated to any objective probability, was first made by Ramsey (1931).
For him, a degree of belief measured the odds at which one would bet on
events, and it was consistency within our betting policy that led it to
obey the probability axioms. Objective reality was not part of the
picture. De Finetti pushed this further (1937), claiming that because
coherent subjective beliefs already reproduce the observed frequencies,
an objective probability standing behind them does no work. He went on
to compare the notion of objective probability to fairies and witches
({de Finetti} 1974).

Such a comparison does not hold up in Davidson's scheme--content dualism
(1974). To call objective probability a fiction, as de Finetti does, is
to say that we looked on the far side of belief and found nothing there.
Such an argument sets up a dualism of belief and the given, with de
Finetti merely evicting objective probability from the given. Davidson,
who rejected such dualisms, would instead deny that there is any
uninterpreted given at all --- what Sellars called the Myth of the Given
(1956). To ask how far \(\mathrm{P}\) has converged on reality is to
picture the belief system on one side and reality --- in the form of
``objective probability'' --- as an uninterpreted given on the other,
i.e.~to accept the third dogma of a scheme--content dualism.

Davidson's case against this dogma is that the uninterpreted given is
not something we can make clear sense of; there is not even a
well-defined distance for our beliefs to close. To see this consider
three different possible meanings for \(p\) that might be possible if
set out to measure the distance between the ``objective'' probability
\(\mathbb{P}(D_3=1)=p\) and the belief
\(\mathrm{P}(D_3=1)=\mathbb{E}[\Theta]\). We might, under the one
meaning, refer to the the actual physical event of the shot. In this
case it is well defined (the shot either goes in or it does not) and it
carries values \(0\) or \(1\), and measuring the distance in this case
is trivial (Hájek 2007). Suppose instead that we fix a reference class,
so that \(p=0.1\), for example. Under this seconding meaning of \(p\) we
have not reached outside the belief system. Comparing
\(\mathbb{E}[\Theta]\) with \(0.1\) is comparing one model with another
--- exactly the operation a belief system is already built to perform.
Finally, a third possibility is that we do not choose a reference class
and \(p\) is undefined, the distance between \(p\) and
\(\mathbb{E}[\Theta]\) is also undefined and there is nothing to
measure. As a result, the question of converging on reality never gets
set up: it fails to pose a task, and the scheme--content gap it
presupposes is, as Davidson says, unintelligible.

Jeffrey took a slightly different approach to de Finetti and argued that
the senses are not ``telegraph lines'' on which the world sends
observation sentences for us to condition upon. He argued that no
inquiry has a certain given to build on, and this led him to be
subjectivist about probability (Jeffrey 1992). Davidson would instead
emphasise the combined process of watching football and building up an
idea of how often goals are scored, as a process of constructing the
telegraph poles. Our experiences give us direct contact with reports on
the game, but these are not separated into one part data and one part
model. We should drop the dualism, Davidson tells us, and thus restore
our unmediated contact with that world.

Jeffrey famously described Bayesian reasoning as ``probabilities all the
way down'' (LaCaze 2006; Jeffrey 2004). Davidson held that ``nothing can
count as a reason for holding a belief except another belief'' (Davidson
1983). But their respective conclusions were very different. While
Jeffrey took ``probabilities all the way down'' to show that probability
is subjective, Davidson took the same absence of a given to show the
opposite: once the dualism of scheme and content is dropped, there is no
gap between belief and world for subjectivity to fall into. The beliefs
we have are already about an objective world. Belief all the way down is
not a veil over reality, but our unmediated contact with it.

The above argument suggests we should keep talking about degrees of
belief when referring to \(\mathrm{P}\), but stop picturing the
probability space as beliefs about a reality lying beyond it. A
probability space is not a screen between us and the world; it is the
very terms in which we are in direct contact with a reality --- our
direct contact with the shots and goals themselves.

\subsection{2.3 Model and data}\label{model-and-data}

The standard Bayesian picture is that model parameters \(\Theta\) are
set against a data set \(D\): the likelihood
\(\mathrm{P}(D=d \mid \Theta=\theta)\) is the probability the model
assigns to the data; the posterior \(\mathrm{P}(\Theta \mid D=d)\) is
our revised belief once the data are in (Gelman et al. 2013; Kass and
Raftery 1995; Bernardo and Smith 1994). Competing accounts of how, for
example, prawns interact, of how democracy develops, or of the shape of
an evolutionary tree give differing sets of model parameters,
\(\Theta\). The same \(D\) is assigned different probabilities under
different parameter values in alternative models. On this picture, the
data and model might be thought to be separate in kind: the data are
given, the model's parameters are fitted to them. Our football example
also appears to have these two parts --- the shots and their outcomes,
\(D_t\), and the goal rate \(\Theta\),. which is a parameter of the
model.

We now consider more carefully what we really mean by data and parameter
in this setting. Our estimate of the goal rate is a statistic of the
counts: the prediction of (2),
\[\mathrm{P}(D_{t+1}=1 \mid G_t) = \frac{G_t}{t} = \frac{D_1 + \dots + D_t}{t},\]
is built from the observations alone. \(G_t/t\) is the average goals per
shot and is, as such, also data. We could have simply kept a running
average of goals per shot and this would have been data on which to fit
our model.

The parameter \(\Theta\) is the goal rate and is also a random variable.
It is tempting to picture \(\Theta\) as some fact lying behind the
counts. But it is just the value those counts settle on: under the
exchangeability we have assumed, de Finetti's theorem identifies it with
the limiting frequency, \(\Theta = \lim_{t\to\infty} G_t/t\). The
estimate and the parameter are then the same kind of thing --- a
finite-sample statistic and its infinite-sample limit, both random
variables on \(\Omega\). The posterior \(\mathrm{P}(\Theta \mid D)\) is
simply our belief about where that limit lies given the counts so far.
The parameter, \(\Theta\), is therefore not a further ingredient
standing behind the data; it is a feature of the data sequence itself.
The fitted model thus carries no content the data did not already hold,
only the structure --- the frame --- within which the counts are read.
\(D_t\), \(G_t\) and \(\Theta\) are at bottom the same kind of thing,
different representations (in the form of random variables) of data.

The same point holds for more complex models. For example, the slope of
a regression line is \(\hat\beta = S_{xy}/S_{xx}\). What is imposed here
is the decision to fit a line, and to make \(y\) the response variable.
Similarly, if instead of updating a parameter \(\Theta\) we are
selecting among models \(M\), then competing models can be gathered into
a single encompassing model carrying an extra parameter that records
which is in force (Kass and Raftery 1995); Bayes' theorem sees no
difference between the two. What we call the model and what we call its
parameters is a decision about which coordinates of one belief system we
hold fixed and which we let vary, not a division the mathematics
enforces.

The inseparability of data and model can also be considered from the
point of view of what data we collect Tulodziecki (2025). In the above
example, before a single number is written down we have decided just to
count shots and goals, not a shot's distance, the state of the pitch,
which goalkeeper it beat etc. By fixing \(\mathrm{P}\) we have already
assumed the shots are independent. If we changed what counts as an
outcome in our model --- record the distance, drop the independence ---
and the ``data'' changes too. What presents itself as the given is the
residue of modelling choices already made. Again, this is not a quirk of
shots example but the ordinary condition of scientific data, which are
corrected, interpolated and filtered through models long before they
reach the analyst who calls them data (Bokulich 2020).

Model and data come apart no more cleanly than Davidson's scheme and
content (1974). They are not two independently given things a framework
brings into contact, but distinctions made within one structure. This
goes further than the pragmatic-representational view of Bokulich and
Parker (2021), on which data are representations to be assessed for
adequacy to a purpose and so remain a category of their own.

\subsection{2.4 A belief system, not an update
rule}\label{a-belief-system-not-an-update-rule}

Much of the appeal of Bayesian reasoning is presented as a story about
updating. We begin with a prior, the data arrive, and we move to a
posterior. But once the probability space is laid down, we see above
that Bayes' theorem is a relation internal to \(\mathrm{P}\). Bayes'
theorem
\[\mathrm{P}(\Theta=\theta \mid D_{t+1}=1, G_t=g) = \frac{\mathrm{P}(D_{t+1}=1 \mid \Theta=\theta)\,\mathrm{P}(\Theta=\theta \mid G_t=g)}{\mathrm{P}(D_{t+1}=1 \mid G_t=g)},\]
is not a bridge between a model and some data lying outside it. It is a
rearrangement of relations that were already fixed the moment we wrote
down \(\mathrm{P}\). Every term in it, \(\Theta\), \(D_{t+1}\) and
\(G_t\) alike, is a random variable within the imposed structure, and
the equation simply describes how those variables stand to one another
within it.

So Bayes' theorem cannot be the place where data meet a model, because
it is a statement wholly internal to the model --- a description of a
relationship we ourselves put there. To call the move from prior to
posterior an ``updating of the model by the data'' is therefore already
to misdescribe it: what looks like data acting on a model is only the
model's own structure, re-expressed. Everything of substance --- the
outcomes we recognise, the events we distinguish, the conditional
independences we assume, the shape of the priors --- is fixed in
\(\mathrm{P}\) before a single shot is taken.

This is clearest if we lay \(\mathrm{P}\) out over models and data
together. Then the prior \(\mathrm{P}(M)\) and the posterior
\(\mathrm{P}(M \mid D)\) are not two successive states of belief but two
features of one and the same function. Observing \(D\) does not alter
\(\mathrm{P}\); it tells us which conditional of \(\mathrm{P}\) to read.
The belief system was already committed, in advance, to every posterior
it would ever report ({de Finetti} 1937). So the data do not change the
belief system as part of the updating process --- they select within it.
Genuine learning happens at another level: when a model fails its checks
and we abandon \(\mathrm{P}\) for a different one. What is internal to
\(\mathrm{P}\) is the conditioning; what is not is the building and
rebuilding of the very \(\mathrm{P}\) we condition within. Gelman and
Shalizi make the same point from the viewpoint of statistical practice.
Rather than counting on Bayes theorem they see the work as proposing a
belief system, checking it against the data, and rebuilding it when it
fails (Gelman and Shalizi 2013).

\section{3. An everyday example}\label{an-everyday-example}

Let's illustrate some of the points we have met so far with an example.

Every Saturday, Joe goes to watch his favourite football team. Since he
was a boy he has kept the same habit: on a slip of paper he writes down,
shot by shot, whether each attempt was a goal or a miss. By the final
whistle he has the whole sequence --- every shot, and which of them went
in. In mathematical terms, Joe has the whole ordered sequence of
outcomes \(D_1, D_2, \dots, D_n\), where \(D_t = 1\) records a goal on
shot \(t\) and \(D_t = 0\) a miss.

His daughter Jen doesn't go to the matches, but she follows them on her
phone. Her app says nothing about the shots that miss; it pings only
when a goal is scored, and shows her the running score and the number of
shots taken so far. So Jen never sees the misses one by one, but she
always knows how many goals have been scored and out of how many
attempts. In mathematical terms, Jen has only the running counts --- the
number of goals \(G_t\) out of the number of shots \(t\) --- and never
the individual outcomes \(D_t\).

Joe's wife Jane has no interest in football and never watches. But she
has heard Joe say the same thing for twenty years: ``this lot score
about one shot in ten''. She has no idea for any specific match how many
shots have been taken, or how many have gone in. In mathematical terms,
Jane has neither the outcomes \(D_t\) nor the counts \(G_t\), only her
standing belief about \(\Theta\), which has a probability distribution
concentrated near \(1/10\).

Let's now look at Joe, Jen and Jane in the context of our points above
about degrees of belief, models and data and belief systems. If all
three of them go to the next match (Joe finally drags Jane to a game),
before a ball is kicked, all three would give the same answer when asked
how likely the first shot is to be a goal: about one in ten. That number
is a belief, not a property of the shot (which might turn out to be an
early penalty or a long distance effort that was never going to go in).
All three of them have, in different ways, formed beliefs that
correspond with reality.

They each hold different data. Joe has a full sequence of shots, Jen has
two running totals and Jane has a single rate parameter. If so inclined,
both Joe and Jen could use their data to construct their own estimate of
the probability distribution of \(\Theta\). But for Jane the 1 in 10 is
the data. Similarly, Jen's counts are Joe's sequence with the order
thrown away. One person's data is another person's model and vice versa.
Across the three of them, we can't make a clean separation of model and
data.

Beneath the different representations, the three do \emph{not} share one
single belief system. Nor have any of them reflected on the axioms of
probability that are so closely related to all three of their individual
belief systems. It is this agreement between four distinct belief
systems --- those of Joe, Jen and Jane, together with the theoretical
one written in the language of probability --- not a single system held
in common, that provides shared predictions about reality. We can
readily translate between the systems --- Joe screaming ``yes!!!'';
Jen's phone lighting up in the ground when the team score; Jane finally
experiencing her husband's excitement; the incremental parameter update
in a Bayesian representation in computer code --- but they are not the
same thing.

After the match, something is bothering Jen. She noticed that after the
first goal, the striker tried another shot, from distance and missed
terribly. The same thing after the second goal, and the third\ldots{} It
made her think: do teams miss more often directly after they have
scored? Could she be missing something by not going to the match (over
and above seeing her Dad's excitement)? When she got back she went
through her Dad's old data. She calculated
\(\mathrm{P}(D_{t+1}=1 \mid  G_t=1)\) and
\(\mathrm{P}(D_{t+1}=1 \mid G_t=0)\), and found they were different. Not
wildly different: the first was \(0.0955\), the second was \(0.1005\),
but all the data her dad had collected allowed her to see that a team
was slightly less likely to score directly after it had just scored. The
average rate of scoring was the same as they had previously thought,
\(0.1\), but she had revealed a new, hidden relationship in the data.

In making her discovery, Jen went directly from examining the
conditional probabilities to identifying a relationship no one in her
family had thought of before. From her finding we could create a new
mathematical model, with an additional parameter to weigh the last shot
contribution, and compare Bayes factor of two models. This would support
her finding, but it is not the root of Jen's discovery. She updated her
belief system as a result of learning something new.

\section{4. Coherence is not the only
constraint}\label{coherence-is-not-the-only-constraint}

Ramsey, de Finetti, Savage and Jeffrey all emphasised that a rational
belief system should be coherent: its degrees of belief must obey the
axioms of probability and thus not lead someone using it to make
inconsistent bets (Ramsey 1931; {de Finetti} 1937; Savage 1954).
Coherence is a constraint from within, governing how a belief system
hangs together. Coherence in this sense --- conformity to the
probability axioms --- is not the graded notion of the
confirmation-theoretic literature (Douven and Meijs 2007), nor
Davidson's own, on which coherence is ``nothing but consistency''
(2006).

Davidson agreed that a belief system must be coherent, though (as we saw
earlier) he denied that this made it subjective. He further argued that
coherence is not enough. Setting out the unified theory, he put it this
way: ``the policy is to assume the speaker's beliefs are logically
consistent. Logical consistency insures no more than the interpretation
of the logical constants, however \ldots{} Further interpretation
requires the assumption of further agreement between speaker and
interpreter. The assumption is certainly justified, the alternative
being that the interpreter finds the speaker unintelligible'' (Davidson
1980). For something to be a belief system at all --- a set of states
with determinate content, each about something and capable of being true
or false --- it must also be interpretable. A belief system is
interpretable when another person could, from someone's words and
actions, make sense of it as a set of beliefs about a shared world. Jane
can read Joe's mood when he comes home as beliefs about the match, and
determine what he believes. In contrast, if Jane just found a notebook
of pages of internally consistent calculations of \(D_t\), \(G_t\) and
the probability distribution of \(\theta\) without any reference to
football, she would not be able to ascertain anything about Joe's
beliefs about the game.

Interpretability, Davidson held, imposes two requirements that coherence
alone does not. The first is charity: an interpreter can fix what a
believer's states mean only by taking them to be largely true, and
largely in agreement with the interpreter's own (Davidson 1974, 1983).
When Jane listens to Joe talking about the match, she assumes he is
telling the truth about the shots and and the frequency of goals. If
Jane thought Joe made up what he had seen then there would be nothing
determinate for the beliefs to be about. The second is triangulation:
what a belief is about is fixed only through the interaction of two or
more agents responding to a common cause in a shared world (Davidson
1991, 1995). Joe and Jen both receive reports from the same match,
albeit in very different ways, giving them a shared sense of what is
correct about the match. A belief system is thus constrained not only
from within, by coherence, but from without, by charity and
triangulation --- by what it takes to be a system of beliefs about an
objective world at all.

The importance of charity and triangulation constraints can be
illustrated by considering prior distributions. For the subjective
tradition, priors for models are unconstrained, because they do not
impact coherence. Indeed, a subjective probability may be almost
anything at all so long as it is coherent (Norton 2008). Jeffrey tried
to address this worry by stressing that ``\,`subjective' probability is
not something fetched out of the sky on a whim; it is your actual
judgment, normally representing what you think your judgment should be,
in view of your information to date and of your sense of other people's
information \ldots{}'' (Jeffrey 2004). Following this line, researchers
have developed a range of approaches for building a prior in practice.
In cases where data on related quantities are plentiful, this data can
be used to inform the prior directly (Efron 2010; Gelman and Hill 2007).
Where expert judgement exists, it can be combined in, for example, The
Sheffield framework, which asks for a full distribution rather than a
single number and guards against biases (Garthwaite et al. 2005; O'Hagan
et al. 2006). And failing the existence of data or experts we can use
weakly informative priors that rule out absurd values while remaining
cautious about what we know (Gelman 2006; Gabry et al. 2019; Gelman et
al. 2020).

These three approaches --- data pooling, expert judgement and caution
--- can be framed in terms of charity and triangulation. Expert
elicitation is charity made into procedure: to form a prior by pooling
what informed others believe is to fix one's own degrees of belief by
taking other interpreters to be largely reasonable and largely right.
Data pooling is triangulation made into procedure: by setting priors
from the frequencies of related cases we anchor belief to a shared
world. And caution answers to both charity and triangulation at once: a
weakly informative prior, checked by simulating its consequences, is
tested before any data are seen against what a competent observer would
find a plausible range of outcomes. Each of the methods for setting
priors thus has a rational motivation, in terms of a requirement to make
our belief system interpretable.

While priors serve as an example, the constraints of charity and
triangulation apply to our entire belief system: from what to include in
a model to how it is parameterised. Which events to admit into
\(\Omega\), whether a shot counts only as goal-or-not or is graded by
distance, whether shots are treated as independent or exchangeable,
which rival models are proposed --- each is a decision about the belief
system. And each is disciplined in the same way as the prior: not by
coherence but by charity and triangulation. A modeller who individuated
goals in a way no colleague could recognise, or framed the game in terms
no fellow analyst could follow, would not have a bolder model but an
uninterpretable one; and a framing answers to the shared world through
the same practices --- replication, peer review, comparison with how
others model the same phenomenon, and the confrontation of the model
with the data read another way. What present themselves as the analyst's
free stipulations are thus constrained throughout by the demand that the
belief system stay interpretable as beliefs about a world held in
common.

The superforecasters of the Good Judgment Project calibrated their
probabilities against others and updated their beliefs in small, shared
steps rather than idiosyncratic leaps (Tetlock and Gardner 2015; Mellers
et al. 2015). That the most accurate reasoners were also the most
thoroughly triangulated is just what the present account predicts: a
belief system improves less by tightening its internal consistency than
by staying answerable to a world held in common.

The building, checking and rebuilding of models against a shared world
that Gelman and Shalizi describe (2013) are methods of triangulation. In
their article, Gelman and Shalizi argue that Bayesian practice is not
the inductive accumulation of belief but a hypothetico-deductive cycle
--- build a model, deduce its consequences, check them against the data,
rebuild when the check fails --- and they align this with Popper's
falsification and with Mayo's severe testing (1959; 2018). Falsification
is only half the story: it tells us when to abandon a belief system, but
not how to build a new one. Charity and triangulation can help here, by
insisting that the construction and revision of a model is disciplined
by answerability to other interpreters and to a shared world. The unit
that is built and revised is the whole belief system, not the isolated
conjecture.

\section{5. The model is not wrong}\label{the-model-is-not-wrong}

Another consequence of Davidson's radical interpretation is that our
mathematical \(\mathrm{P}\) should not be thought of as standing over,
for example, Joe, Jen and Jane. It is just one more belief system, held
by one more believer (the mathematician), it enters the same
triangulation as the others and it is interpreted under the same charity
by others. It is not a master copy (Davidson 1986). A model can be more
explicit, more coherent, wider in reach and better calibrated over many
seasons than anything Joe carries in his head, and where it is, we
should prefer it. But the preference is earned through coherence,
demonstratable charity and triangulation --- and not conferred by the
fact it is expressed in equations.

On a similar note, radical interpretation tells us to (charitably)
assess the use of ``objective'' rules for prior construction, such as
Jeffreys' (1961) invariance rule (not to be confused with Jeffrey) and
Jaynes' maximum entropy principle (Jaynes 2003), as conventions within a
community. Read this way, these rules do not deliver the one correct
prior read off the structure of a problem; they provide a shared
standard that makes one analyst's credences legible and comparable to
another's --- a device for triangulation.

Charity contrasts with a view, I sometimes hear expressed by practising
statisticians, that ``ultimately we can't know if anything is true''.
This sceptical view, which falls into the trap of the third dogma, sees
statistics primarily in terms of preventing mistakes caused by lack of
rigour or careless interpretation of results. Such reasoning should be
turned on its head. On Davidson's account of interpretation, any
expressed belief system is by its nature mostly true --- not because
error is impossible, but because only against a background of largely
correct belief could an interpreter fix what the words mean at all; an
attribution of wholesale error would leave nothing determinate to be
wrong about (Davidson 1995).

We do occasionally, like Jen in the story above, find an error in
someone elses thinking. But in this case we use our shared truths to
communicate that error clearly. On telling Joe of her finding, he would
say freely admit she was right, using his own (slightly incorrect)
belief system as a scaffold for revising his own belief system. Rather
than helping researchers be ``less wrong'', statistics should help us be
``slightly more right''.

We can put this another way by considering the commonly used phrase
``All models are wrong, but some of them are useful'', which was
originally coined by George Box (1979). He was, when he made this point,
warning practitioners not to chase a ``correct'' model through endless
elaboration, and to judge a model by whether it is importantly wrong for
the purpose at hand (Box 1976). This latter point I agree with. The
bigger the correction the better. But I would rather say ``Most models
are correct, that's why they are useful.'' By this I do not mean that a
model is complete or final, but that a model in serious use is one
already made largely true of the world it addresses --- answerable to
other analysts and to the data through the charity and triangulation
described above. Its usefulness is not a consolation for its falsity but
a measure of how far it has been brought into agreement with a shared
world. Box's warning against chasing a ``correct'' model through endless
elaboration still stands; what I resist is the stronger thought that
being wrong is a model's natural and permanent condition. Our task is to
make models truer by making them more useful, and more useful by making
them truer.

\section{6. Conclusion}\label{conclusion}

Davidson's work, which is now over 50 years old, was originally aimed at
conceptual relativism, which underlay much of the thinking in the
1960s--70s: from Kuhn's incommensurable paradigms, Whorf's linguistic
relativity, and Feyerabend's proposal that there could be radically
different, mutually untranslatable conceptual schemes for understanding
the world. Considered as a substantial development in philosophy at the
time, it showed that the problems raised by relativism (e.g.~``everyone
has their own private truth about the world'') could be dissolved, while
avoiding the need to invoke a ``given'' objective reality.

The same dualist structure recurs throughout the foundational debates of
statistics. Rather than reopen these debates, I have asked what
Davidson's radical interpretation says about probabilistic belief
systems. Its conclusions coincide with much of what a modern Bayesian
already does --- eliciting and pooling expert judgement, borrowing
strength across related cases, and adopting weakly informative priors
checked against their consequences (Garthwaite et al. 2005; Gelman and
Hill 2007; Gabry et al. 2019). What the Davidsonian view shows is that
these are not ad hoc fixes, but consequences of charity and
triangulation, and that the real work of inference is not the updating
of a fixed model but the building, checking and rebuilding of whole
belief systems (Gelman and Shalizi 2013). This approach is similar to
Longino's, and to the social senses in Douglas's taxonomy of objectivity
(1990; 2004), but I do not treat objectivity as something a community
achieves. Charity and triangulation are conditions on there being
determinate beliefs to be objective about, and so hold of any belief
system at all, well or badly run.

The Davidsonian view also allows us to agree (just like Joe, Jen and
Jane), even when we are not using exactly the same model. The point of
inference is not to be less wrong against a reality we can never reach,
but to build belief systems that answer ever more fully to one another
and to the world. Most models, in that sense, are already correct; our
task is to make them more so.

\section{References}\label{references}

\protect\phantomsection\label{refs}

\end{document}